\documentclass[12pt]{article}

\usepackage{authblk}
\usepackage{parskip}
\usepackage[normalem]{ulem}

\usepackage{fullpage}
\usepackage{url}
\usepackage{verbatim}
\usepackage{amstext}
\usepackage{amsmath}
\usepackage{amsfonts}
\usepackage{mathtools}
\usepackage{float}

\usepackage{amsthm}
\newtheorem{theorem}{Theorem}[section]

\newtheorem{corollary}[theorem]{Corollary}
\newtheorem{lemma}[theorem]{Lemma}

\newtheorem{claim}[theorem]{Claim}
\newcommand{\vm}[1]{| {#1} \rangle}
\usepackage{hyperref}
\usepackage{color}

\newcommand{\bC}{\mathbb C}
\newcommand{\fS}{\mathfrak S}
\newcommand{\fSZ}{\mathfrak S_{\mathfrak Z}}

\newcommand{\tfS}{\widetilde{\mathfrak S}}
\newcommand{\fA}{\mathfrak A} 
\newcommand{\re}{\mathbb R}
\newcommand{\csg}{{\mathbb C}\mathrm{SG}}
\newcommand{\cyl}[1]{\mathrm{cyl}(#1)}
\newcommand{\mx}{\mathrm{max}}
\newcommand{\mn}{\mathrm{min}}
\newcommand{\fZ}{\mathfrak{Z}}
\newcommand{\fC}{\mathfrak{C}}
\newcommand{\cZ}{\mathcal Z}
\newcommand{\muv}{\mu_\mathbf{v}}

\newcommand{\cH}{\mathcal H}
\newcommand{\card}{\mathfrak N}
\newcommand{\cP}{\mathcal P}
 
\newcommand{\bD}{\bar D} 
\newcommand{\ket}[1]{| #1 \rangle}

\newcommand{\braket}[2]{\langle #1| #2\rangle}
\newcommand{\hO}{\widehat O}
\newcommand{\var}[1]{| #1 |}
\newcommand{\norm}[1]{|| #1 ||}

\newcommand{\zin}{\zeta_n^i}
\newcommand{\zmx}{\zeta_n^\mx}
\newcommand{\zmn}{\zeta_n^\mn}
\newcommand{\ind}{\mathfrak {I}} 
\newcommand{\zmxr}{\zeta_r^\mx}
\newcommand{\zmnr}{\zeta_r^\mn}
\newcommand{\cT}{\mathcal{T}}

\newcommand{\stem}{\mathrm{stem}}
\newcommand{\cI}{\mathcal I}
\newcommand{\hq}{\hat q}
\newcommand{\cO}{\mathcal O}
\newcommand{\bP}{\mathbb P} 
\newcommand{\cN}{\mathcal N} 

\usepackage[margin=1.0in]{geometry}

\usepackage{bbold} 
\newcommand{\one}{\mathbb{1}}
\title{ A Criterion for Covariance \\ in Complex Sequential Growth Models}
\author[1]{Sumati Surya}
\author[2]{Stav Zalel}
\affil[1]{Raman Research Institute, CV Raman Ave, Sadashivanagar, Bangalore, 560080, India.}
\affil[2]{Blackett Laboratory, Imperial College, London, SW7 2AZ, U.K.}
\begin{document}

\maketitle
\abstract{The classical sequential growth model for causal sets provides a template for the 
  dynamics in the deep quantum regime. This growth dynamics is intrinsically temporal and causal, with each new
  element being added to the existing causal set without disturbing its past.  In the quantum version, the probability
  measure on the event algebra is replaced by a  quantum measure, which is Hilbert space valued.  Because of the temporality of the
  growth process, in this approach,  covariant observables (or beables) are measurable only if the quantum measure extends to the
  associated  sigma algebra of events. This is not always guaranteed.  In this work we find a
  criterion for extension (and thence covariance)  in  complex
  sequential growth models for causal sets. We  find a large family of models  in which the measure  extends, so that {\it all} covariant observables are
  measurable. }

\tableofcontents
\section{Introduction}
\label{intro.sec}

One of the most challenging quests in any approach to non-perturbative quantum gravity is in finding a consistent dynamics for the
full  theory.  Within each approach the formulation of the dynamics acquires specific features, not all of which can be
translated to other approaches. In causal set quantum gravity \cite{Bombelli:1987aa},  the emphasis is on the space of
discrete histories or causal sets, with  the dynamics given by a Hilbert space valued measure or equivalently a decoherence functional.  As in  the continuum path
integral, where  each (fixed dimensional)  Lorentzian spacetime appears with a  complex weight, in causal set theory
(CST)  each countable causal set appears in the  path sum with a complex weight.  In continuum-inspired models, the
measure is given in terms of the discrete Einstein-Hilbert or Benincasa-Dowker action
\cite{carliploomis,2dorders,hh,onset,fss,dimred}, but this is not the most natural choice from a fundamental,  order theoretic
perspective. 

One such ``bottom-up'' approach to CST dynamics is the sequential growth paradigm, the classical version of which serves as a
template for the quantum dynamics \cite{Rideout:1999ub,evidence,Martin:2000js}. In this paradigm, the causal set is grown element by element, starting with an
initial element. At every stage of the growth the new element can be added to the future of an existing element or left 
unrelated to it, with some transition probability or amplitude (depending on the case at hand), so that  the past of the existing elements is not changed.  In the classical growth models, this generates a probability measure space $(\Omega, \fZ, \mu)$
where $\Omega$ is the space of all past finite {\it labelled} causal sets, $\fZ$ is {an event algebra (or collection
of all measurable sets) closed under finite set operations} over $\Omega$ and $\mu$ is a probability measure. 

Requiring the dynamics to be Markovian, covariant (path
independent) and causal,  reduces the space of possible probability measures drastically, each characterised by a single transition
probability per stage of the growth \cite{Rideout:1999ub}.  While these probabilities themselves are covariant, the events in $\fA$ are
not, since they are generated by finite stage events in $\Omega$.  Covariant events (which are the ``beables'' of this
theory and which we will sometimes refer to as  covariant 
observables),  can only be
defined after generating the infinite stage events.  This means that  in order to construct {\it all} possible covariant
events from $\fZ$, one has to go to the full sigma-algebra $\fSZ$ generated by $\fZ$. The covariant events are given by the
quotient-sigma-algebra $\tfS =\fSZ/\sim$ where the equivalence relation $\sim$ is over relabelings of causal sets in
$\Omega$ \cite{Brightwell:2002yu}\footnote{A formulation of the growth dynamics generated by
 covariant events to was adopted in \cite{dz} using ``stem
  events''. However, we will not pursue this approach  here.}.  Because $\mu$ is a probability  measure, by the
Kolmogorov-Caratheodary-Hahn extension theorem \cite{Kolmogorov:1975}, it possesses a unique extension  to $\fSZ$ and hence 
one can in principle calculate  the measure of covariant events.  Examples  of covariant events are (a) the ``originary'' event which is the
collection of causal sets with a  single element to the past of all other elements: this is the  analogue of a ``big bang'' (b) the
post event which is the collection of histories each  containing at least one element such that all other elements are
either to its past or its future: this is the analogue of a ``bounce''.  

In quantum sequential growth  models, the idea is to  replace the probability measure by a  ``quantum measure'', which can
be realised as  a finitely additive  vector measure $\muv$  valued in a   ``histories''  Hilbert space $\cH$
\cite{hilbert,djs}. As in the classical growth models, the quantum dynamics  is then characterised by the  quantum triple
$(\Omega,\fZ, \muv )$.  The simplest quantum version of the growth models is obtained by complexifying 
the  classical probability measure, so that $\muv$ is valued in $\bC$. This is the Complex Sequential Growth or
$\bC$SG dynamics that is the focus of this present work.

Such a simplification does not however guarantee the extension of
$\muv$  to  the full sigma algebra $\fSZ$; it must additionally satisfy  certain boundedness conditions 
\cite{diesteluhl}. As shown in \cite{djs},  for complex percolation ($\bC$P), where
the dynamics is characterised by a single  complex number $q$, the measure  does {\it not}  extend and hence cannot be
defined for covariant events, unless $q\in [0,1]$, i.e., for ``real''  $\bC$P  ($\re\bC$P). While the latter  is not in itself
strictly classical,  it is a fairly trivial example of $\bC$SG. It is therefore  of interest to to find a larger class
of  $\bC$SG models  in which $\muv$  can be extended to $\fSZ$. 

In \cite{sec} it was argued that not all covariant events may be physically relevant
and that it would be sufficient for the measure to extend to a subclass of covariant events via  {some {\it {
    conditional}}}  
convergence conditions. It can be shown that one such condition is satisfied by the measure of  the originary
  event  in the $\bC$P model \cite{ss-ec}.  However, apart from a   
simple  class of covariant events, which includes the originary event, setting up a conditional convergence protocol for
other  covariant events  like the post event becomes rapidly more cumbersome.  It is therefore desirable to look for quantum measures
$\muv$ that extend to the full-sigma algebra $\fSZ$, so that  {\it every } covariant event is  measurable.  Such
models thus define a consistent covariant dynamics.  

In this work we find a criteria for $\muv$ to extend to $\fSZ$ in $\bC$SG models.  We find by explicit construction
large classes of $\bC$SG models that admit an extension and hence define consistent covariant dynamics, as well as those which
do not.  Our methods follow the spirit of the analysis of the $\bC$P dynamics in \cite{djs}, where the extension of the
measure is related to a colinearity criterion.  

In Section \ref{gp.sec} we review the sequential growth paradigm, where we define the event algebra $\fZ$ generated from
finite labelled causal sets and the associated {\sl cylinder sets} in $\Omega$.  We then review the  CSG 
models of \cite{Rideout:1999ub,Rideout:2001kv} in Section \ref{csg.sec} which serve as a template for the quantum
dynamics. Next,  we define
QSG models broadly and  the subclass of $\bC$SG dynamics  in Section \ref{qsg.sec}. In Section
\ref{extension.sec} we use a distilled version of the  
Caratheodary-Hahn-Kluvnek(CHK)  theorem for complex measures on $\fZ$ (proved in Appendix \ref{vm.app}), which states
that bounded variation is a necessary and sufficient condition for the extension of $\fZ$ to $\fSZ$.  Section
\ref{results.sec} contains our main results. In Section \ref{criteria.sec} we find criteria for bounded variation,
summarised in Theorem \ref{one.thm}.  In Section \ref{complex.sec} we translate these criteria  to the specific case of
$\bC$SG  by proving two Lemmas \ref{amax.lemma} and \ref{amin.lemma} which gives us a useful Corollary \ref{zetaca.cor}
to Theorem \ref{one.thm}.  Finally in  Section \ref{examples.sec}
we give explicit examples of  $\bC$SG models that extend and some that do not. In Section \ref{discussion.sec} we
discuss  how these results can be used to make predictive statements about covariant observables in quantum gravity.
Appendix \ref{defs.app} lists a few of the standard definitions from 
causal set theory. The list is not exhaustive and we refer the reader to the literature
\cite{Rideout:1999ub,livingreviews}.   In Appendix \ref{vm.app} we show how the CHK theorem
implies Theorem \ref{complexCHK.thm} for a complex measure  over $\fZ$.

\section{The Sequential Growth Paradigm}
\label{gp.sec}

In CST there is a natural correspondence between the cardinality $n$ of spacetime regions and the continuum spacetime volume. In the unimodular
approach to gravity, the latter appears as a natural ``time-parameter''. Hence evolution corresponds to increasing
spacetime volume (normalised appropriately). This translates in CST to an increase in the cardinality of the causal
set so that the causal set ``grows'' element by element. This motivation is at the heart of the sequential growth paradigm. 

\begin{figure}[htpb]
  \centering
	\includegraphics[width=0.60\textwidth]{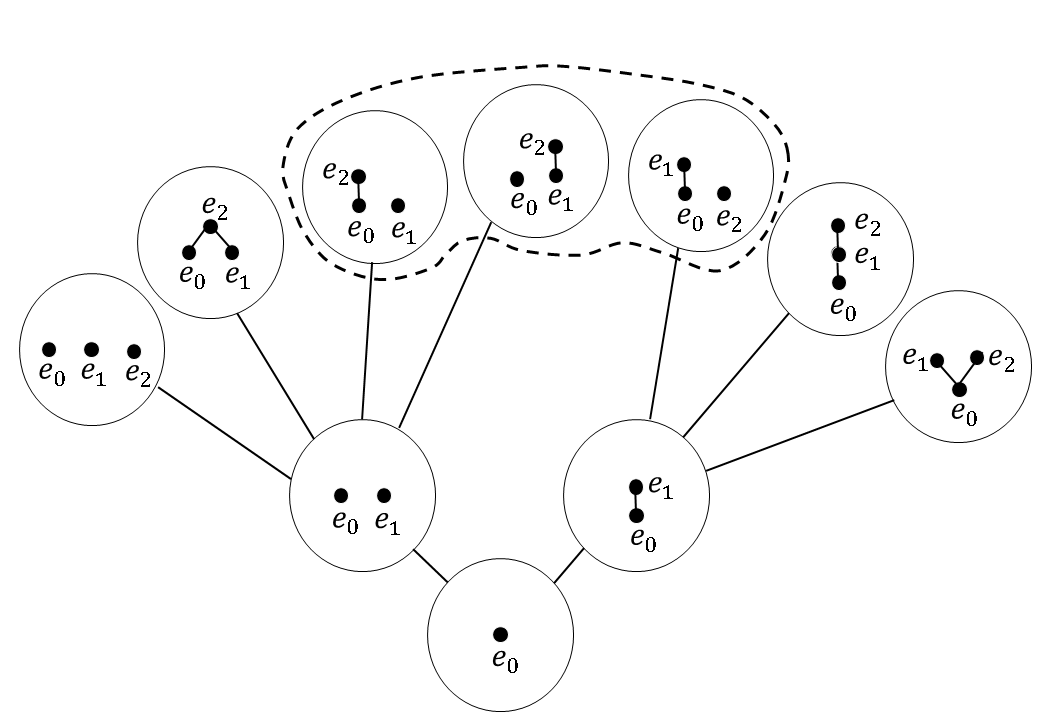}
	\caption{First three stages of sequential growth. The 3-element causal sets that are order-isomorphic to each
          other are marked.}
	\label{poscau.fig} 
\end{figure}

A natural starting point for the growth process is therefore at $n=1$, where, with certainty,  a single element $e_1$ is
born.  At stage $n=2$, the new element $e_2$ can be  added either to the future of $e_1$ to form a $2$-element chain, or left  unrelated
to it, to form a $2$-element anti-chain\footnote{See Appendix~\ref{defs.app} for basic CST definitions.}. However, it cannot be added to the past of $e_1$.  At every stage $n$, the new
causal set element $e_{n+1}$ is ``added'' to the existing causal set $c_n$  so that it is either  to the future of some of the
elements or left unrelated to them. Importantly, it does not change the past of any of the elements in
$c_n$\cite{Rideout:1999ub}.
Fig.~\ref{poscau.fig} is an
illustration of this process upto stage $n=3$. 
In \cite{Rideout:1999ub} this is referred to as {\sl internal temporality}. This condition  is independent of the choice of the
measure, and defines a  growth poset  or  tree  of labelled causal sets, termed {\sl poscau} $\cP$.  We will refer to
each finite labelled poset in the tree as a {\sl node}. The (unique) set of nodes from $e_0$ to an $n$-element node
will be referred to as the {\sl $n+1$-jointed  branch} associated with this node.     

As $n \rightarrow \infty$, this growth process generates the sample space $\Omega$ of countable {\it labelled} past finite causal sets.
The labelling  is evident from Fig.~\ref{poscau.fig}, which shows that in some instances the new element at stage $n$ could
have been added at an earlier stage to get the same {\sl unlabelled}  causal set at stage $n$. As an example, consider the
three labelled $n=3$-element causal sets marked in Fig.~\ref{poscau.fig}.   These are all the 
same unlabelled causal set, but with different time labels corresponding to how they were created.  (i) At stage $n=1$ the element $e_1$ is either unrelated to $e_0$ (in the
left two cases) or is to its future (in the third case) (ii) At stage $n=2$ the element $e_2$ is added to the
future of either  $e_0$ or $e_1$ giving rise to the two figures on the left, or is unrelated to them as in the third
figure.  Again, what is evident is that the labelling must satisfy the order relation $e_i \prec e_j \implies
i<j$. This is referred to as a {\sl natural labelling} or a  {\sl linear extension}. We will henceforth call two distinct labelled
  causal sets $c,c'$  {\sl
  order-isomorphic} to each other (denoted by $c\sim c'$) if they are labelings  of the same
unlabelled causal set. We refer the reader to the literature \cite{Rideout:1999ub,Brightwell:2002yu,dz} for a more
detailed discussion of this terminology. 

Next, one must define the {\sl measurable sets} which constitute  the {\sl event algebra}, which is a  field of subsets of $\Omega$  closed under finite set
operations and includes $\Omega$ and $\emptyset$.
The event algebra naturally associated with the above  growth process is generated by the nodes in $\cP$.  Let
$\Omega_n$ denote the set of $n$-element labelled causal sets, which is of finite cardinality $\card_n\equiv |\Omega_n|$ for finite $n$. For example, using Fig.~\ref{poscau.fig}
we find  that $|\Omega_2|=2$ and $
|\Omega_3|=7$, while for  large $n$ the growth is super-exponential, with   $|\Omega_n| \sim
2^{n^2/4}$, to leading order \cite{kr}.  Each finite labelled causal set $c_n^i \in \Omega_n$, $i \in
\ind(n)=\{1,\ldots, \card_n\}$
is a { node}  in $\cP$ and, being labelled,  also represents its history of formation, i.e., the 
unique  $(n+1)$-jointed  branch in $\cP$, starting from $e_0$\footnote{In the $1+1$ random walk on a lattice, this is analogous to a
particular choice of  the $(n+1)$-jointed path $\{x_0, x_1, \ldots, x_n\}$ for an fixed initial location $x_0$.}.   Thus, for   each  { node} $c_n^i$  we
can  associate a  {\sl cylinder set} 
\begin{equation}
  \cyl{c_n^i} \equiv \{c \in \Omega \, \, | \, \, c|_n=c_n^i\}, \, \, \cyl{c_n^i}\subset \Omega
  \end{equation} 
where $c|_n$ denotes the first $n$ elements of the labelled causal set $c\in \Omega$.  Because 
 $\cP$ is a tree, cylinder sets satisfy the {\sl nesting property}  
\begin{equation}
  \cyl{c_m^i} \cap \cyl{c_n^j} \neq 0, \Rightarrow \cyl{c_n^j} \subset \cyl{c_m^i},  \,\, \mathrm{for} \,\,  m<n  .
  \label{nested.eq} 
\end{equation}
In other words, a non-trivial intersection between two distinct cylinder sets is possible only if one is a proper subset of the
other.

Because $\cP$ is a tree, for any $c_n^i \in \Omega_n$,
\begin{equation}
  \cyl{c_n^i} = \bigsqcup_{j(i)} \cyl{c_{n+1}^{j(i)}}.
  \label{defji.eq} 
\end{equation}
where  $\fC(c_n^i) \equiv \{ c_{n+1}^{j(i)} \} $ denotes the  set of  {\sl children}  of $c_n^i$ in $\cP$, i.e. the set of $n+1$ element
causal sets  emanating from the $c_n^i$ node  in $\cP$. We use the functional notation  $j(i)$ to denote that $j$ is  valued
in an index set $\ind(i,n) \subset \ind(n)$  of cardinality
$|\fC(c_n^i)|$, which depends  on $i$, or equivalently,  $c_n^i$.  For example, from  Fig.~\ref{poscau.fig} we see that
the $n=2$ {\sl antichain}  $c_2^a$ has 4 children, while
the the $n=2$ {\sl chain} $c_2^c$ has 3 children. 

Let $\cZ_n$ denote the collection of  cylinder sets at level $n$ and $\cZ$ the collection of all cylinder sets. The event algebra $\fZ$ is then generated by taking finite unions,
intersections and complements of the elements of $\cZ$. The nesting  property, Eqn.~\ref{nested.eq}, then implies that for any $\alpha \in \fZ$, there 
exists a smallest integer $n_\alpha<\infty$ and a subset ${S_\alpha} \subset \{ 1 , \ldots, \card_{n_\alpha}\}$ such that $\alpha={\bigsqcup_{k\in S_\alpha}
\cyl{c_{n_\alpha}^k}}$. We define the {\sl fine
  partition}  of an event $\alpha \in \fZ$ as  $\cN_\alpha=\{ \cyl{c_{n_\alpha}^k} \}$, $k \in S_\alpha$, of $n_\alpha$-element  nodes in $\cP$. 

Our interest is in events that are covariant.  Following \cite{Brightwell:2002yu} we define a  {\sl covariant  set}   
$\alpha \subseteq \Omega$  as 
\begin{equation}  
\alpha=\{c| c'\sim c {\implies} c' \in \alpha \}. 
\end{equation} 
If $\alpha$ belongs to an event algebra, then we call it a  {\sl covariant event}.  In the language of observables, or
beables, we will also refer to these as {\sl covariant observables}. 

Using the nesting property, we see that  no event $\alpha \in \fZ$ can be covariant unless $\alpha=\Omega$.  Consider the fine partition  $\cN_\alpha$ (defined above) for any   $\alpha\subset \Omega$, so that  $\alpha=\bigsqcup_{k\in S_\alpha}\cyl{c_{n_\alpha}^k}$. Let $c_{n_\alpha}^{s}$ be a node in $\cN_\alpha$  with the largest number of minimal elements $m_\alpha$. (i) Assume $n_\alpha>
m_\alpha$, i.e.,  the $n_\alpha$-element antichain $c_{n_\alpha}^a$ does not belong to $\cN_\alpha$. Let  $c_{n_\alpha+1}^{g(s)}$ denote the  {\sl gregarious child}
of  $c_{n_\alpha}^{s}$, i.e., one  in which  the new element  $e_{n_{\alpha}+1}$ is unrelated to all the
elements in $c_{n_\alpha}^{s}$. Thus,  there exists an $(n_\alpha+1)$-element node  $c_{n_\alpha+1}^l \sim c_{n_\alpha+1}^{g(s)}
$ such that the first  $m_\alpha+1$ elements in $c_{n_\alpha+1}^l$ are the antichain $c_{m_\alpha+1}^a$. But
$c_{n_\alpha+1}^l \not\in \cN_\alpha$ since otherwise $m_\alpha$ would not be  the largest number of 
minimal elements for the set of nodes $\cN_\alpha$. This means that for every $c \in \cyl{c_{n_\alpha+1}^{g(s)}}$, there exists an order-isomorphic  $c' \in \cyl{c_{n_\alpha+1}^{l}}$.   Because of the nested property of cylinder sets, while $\cyl{c_{n_\alpha+1}^{g(s)}} \subset \alpha$, 
$\cyl{c_{n_\alpha+1}^l}\not\subset \alpha$, and hence  $ \alpha$ is not covariant. (ii) If   $m_\alpha= n_\alpha$,
$c_{n_\alpha}^a \in \cN_\alpha$.  Let $\cN_\alpha^c$ denote the (non-empty) complement of
$\cN_\alpha$ in the  set of all possible $n_\alpha$ nodes, and $m_\alpha^c$ the
largest number of minimal elements for any node in  $S_\alpha^c$. The argument (i) then tells us
that $\alpha^c\in \fZ$  is not
covariant. Hence $\alpha$ is not covariant.

This means that the event algebra $\fZ$ does not suffice to be able to define covariant observables.  In order to do so, one needs
to include events obtained from  {\it countable}  set operations on $\fZ$.  An example of a covariant event is  the originary
event $\alpha_{\mathrm{orig}}$ (mentioned earlier) where there is a single element to the past of all the 
other elements in the causal set, analogous to a big bang. $\alpha_{\mathrm{orig}}$ is invariant under natural relabellings since the
initial element must always come at stage $n=0$. In the sequential  growth process, at any  finite   stage
$n$,  the gregarious child is not originary and hence every $\cyl{c_n^i} \in \fZ$ contains causal sets that are not
originary, even if $c_n^i$ itself is originary.  However, $\alpha_{\mathrm{orig}}$ can be constructed from {\it
  countable}  set
operations. Its complement, $\alpha_{\mathrm{orig}}^c$, is the union of causal sets
which are non-originary, i.e., causal sets that contain a $2$-element subset $c_2 $ which is its own past,  and such that {$c_2\sim
c_2^a$}, so that 
\begin{equation}
\alpha_{\mathrm{orig}} = \biggl(\bigsqcup_{{n>0}} \bigsqcup_{i \in \cI_n} \cyl{c_n^i}\biggr)^c, 
 \end{equation}  
where $\cI_n$ labels the $n$-element nodes for which the $n^{\mathrm{th}}$ element is the only gregarious element. This
construction is analogous to the one for the  return event in the discrete random walk, which again uses countable
unions of finite time events.

The smallest algebra that includes events generated by countable set operations on $\fZ$ is its associated sigma-algebra
$\fSZ$.  The set of covariant events themselves form a sigma-algebra which is a  sub-sigma-algebra of $\fSZ$
 \cite{Brightwell:2002vw}. Equivalently, one can build covariant events from $\fSZ$ by taking equivalence classes of
 causal sets under relabellings. In the latter approach, if $\sim$ denotes equivalence under relabellings, the
 sigma-algebra of covariant events is the quotient sigma-algebra $\fSZ/\sim$.
 
We note that this is not the only way to construct covariant events. In the approach of  \cite{dz} instead of $\fZ$, one
considers an event algebra that  is generated from covariant ``stem'' events. The dynamics is defined as a random
walk on the associated covariant tree of posets.

\subsection{Classical Sequential Growth}
\label{csg.sec} 

We begin by describing the classical sequential growth process of \cite{Rideout:1999ub}.  The dynamics on $\cP$ is a
specification of the  measure over $\fZ$. As in the random walk, one can assign a measure to $\fZ$ by letting
{$\mu(\cyl{c_n^i}) \equiv \mathbb{P}(c_n^i)$, where $\mathbb{P}(c_n^i)$ is the probability that a directed random walk
  from the origin in $\cP$ reaches the node $c_n^i$ by stage $n$, and is determined by the particular growth process. {This choice of measure ensures that $\mu$ is a finitely additive {\sl
  probability measure}, {i.e.}, $\mu:\fZ  \rightarrow [0,1]$ and $\mu(\Omega)=1$. By the Kolmogorov-Caratheodary-Hahn
extension theorem, $\mu$ extends to $\fSZ$, and hence to the sigma-algebra of covariant events.}

As  discussed in \cite{Rideout:1999ub} there are certain natural conditions to
impose on the measure for the classical sequential growth.  The first is 
{\sl (a) Covariance}, i.e., the measure  is the same  for order-isomorphic causal sets. In Fig
\ref{poscau.fig} there are three  $n=3$-element order-isomorphic causal
sets whose associated cylinder sets must therefore have the same measure. The second is that the transition probabilities satisfy a  {\sl (b) Markovian sum
  rule}  
\begin{equation} 
  \sum_{j(i)}\bP(c_n^i \rightarrow c_n^{j(i)} ) = 1, 
\end{equation} 
where $j(i)$ is valued in an index set $\ind(i,n)$ of cardinality $|\fC(c_n^i)|$, for all nodes in $\cP$\footnote{It is
  important to note that if there are $k$ order-isomorphic children in a given transition, then the measure not only
  counts each equally, but the multiplicity $k$ appears  in the Markovian sum. In this sense the measure does
  not treat order-isomorphism as ``gauge''. }.

Finally, there is the dynamical causality rule which we term {\sl (c)  Spectator Independence}\footnote{This is referred to in
\cite{Rideout:1999ub}  as ``Bell Causality''. The reason to shy away from this terminology in the present work is its implications
for quantum entanglement, which we will not discuss.}, which needs a little more terminology to define. Let {$c_n^i \rightarrow c_{n+1}^{j(i)}$} be a transition and
define the associated  {\sl precursor set} to be the past of  the new element $e_{n+1}$. If the precursor set is all
of $c_n^i$  this transition is described as {\sl timid} and if it is the empty set, it is described as {\sl
  gregarious}, introduced previously.  Those elements in $c_n^i$ not in the  precursor set of $e_{n+1}$ are then termed {\sl spectators}. The
idea of condition (c)  is that the transition cannot depend {\it explicitly}   on the spectators, and is hence intrinsically
causal.

{Consider two non-timid transitions  $c_n^i \rightarrow c_{n+1}^{j_1}$ and $c_n^i \rightarrow c_{n+1}^{j_2}$, with 
$j_{1},j_2 \in \ind(i,n)$, and with spectator sets  $P_{1},P_2$ respectively, and consider an $m$  element causal set, $c_m^k$
in $\cP$, which is order-isomorphic to $P_1\cup P_2$.  Then there exists  children $c_{m+1}^{l_1}, c_{m+1}^{l_2}$   of
$c_m^k$, with $l_1,l_2 \in \ind(k,m)$  such that the precursor set of the new element in $c_{m+1}^{l_1}$ is order-isomorphic
to $P_1$,  and that of the new element in is $c_{m+1}^{l_2}$ is order-isomorphic to $P_2$.}

The requirement (c) can then be expressed as
\begin{equation}
\frac{\bP(c_n^i \rightarrow c_n^{j_1})}{\bP(c_n^i \rightarrow c_n^{j_2})} = \frac{\bP(c_m^k \rightarrow c_m^{l_1})}{\bP(c_m^k \rightarrow c_m^{l_2})} 
  \end{equation} 
 This condition can be reformulated as a product rule, which holds even when some of the
transition probabilities  are set to zero \cite{rv,ds}.

These three conditions on the transition probabilities simplify the dynamics drastically so that at every stage one has
a single independent coupling constant. It is convenient to take this to be the transition
probability  $q_n$ from $c_n^a$ to  $c_{n+1}^a$\cite{Rideout:1999ub}.  For a generic  transition at stage $n$,  $c_n^i \rightarrow
c_{n+1}^{j(i)}$, the transition probability is given by  
\begin{equation}
{\mathbb{P}(c_n^i \rightarrow
c_{n+1}^{j(i)})}= \sum_{k=0}^m(-)^k \binom{m}{k} \frac{q_n}{q_{\varpi-k}}, 
  \end{equation} 
where  $\varpi$ is the cardinality of the precursor set, and $m$ denotes the number of maximal elements in the precursor
set. 
Alternatively, one can use  the coupling constants $t_n$, 
  \begin{equation}
t_n=\sum_{k=0}^n(-)^{n-k}\binom{n}{k} \frac{1}{q_k}.  
  \end{equation} 
in terms of which the transition probabilities are 
\begin{equation}
{\mathbb{P}(c_n^i \rightarrow
c_{n+1}^{j(i)})}=\frac{\lambda(\varpi,m)}{\lambda(n,0)}, \quad \mathrm{ where} \quad \lambda(a,b)=\sum \limits_{k=b}^{a}
\binom{a-b}{k-b} t_k.
\label{alphan.eq}
\end{equation}

One of the simplest growth models is {\sl transitive percolation}, where  $q_n=q^n$ {and $0<q<1$}, or
equivalently $t_n=t^n$ {and $t>0$}, so that
there is a single parameter that governs the growth. One also has the deterministic  {\sl dust universe} with $q_n=1$, or $t_0=1,
t_k=0, k \geq 1 $, so that only the antichain is generated and  the {\sl forest universe}  {in which all transition
  probabilities are equal and are given by $\mathbb{P}(c_n^i \rightarrow c_{n+1}^{j_1})=q_n=(1+n)^{-1}$, or
  equivalently, $t_0=t_1=1$, $t_k=0, k \geq 2$. The forest universe generates, with unit probability, a causal set which
  is tree-like, with each element in the causal set having a single past {\sl link}, which is a relation that
  cannot be inferred from transitivity.}

\subsection{Quantum Sequential Growth}
\label{qsg.sec}

We wish to construct a  quantum dynamics on the tree, $\cP$. {To do so, we will follow the method of \cite{djs,Dowker:2010ng}.} The growth paradigm describes the kinematics, while the dynamics is encoded in the measure.  This means that  both $\Omega$ and $\fZ $,
generated by the collection of cylinder sets $\cZ$ remain as in Section \ref{gp.sec}, but the 
probability measure is replaced by a {\sl quantum measure}, which we define as follows. 
A {\sl quantum measure} is a Hilbert space $\cH$ valued vector measure $\muv$  on an  event  algebra $\fA$ which is finitely 
additive, i.e., for any finite collection of pairwise disjoint events $\{\alpha_i\}$, $\alpha_i\in \fA$,
\begin{equation}
\ket{\sqcup_i\alpha_i}=\sum_i \ket{\alpha_i}
\end{equation}
where $\ket{\alpha}\equiv \muv(\alpha)$. If $\fA$ is also a sigma algebra, then the vector measure is also required to be 
countably additive.  
In either case, the norm  {squared} of  $\muv$ is not additive (finitely or countably, as the case may be) since  in general 
\begin{equation}
{\braket{\alpha}{\alpha} = \braket{\sqcup_j\,\alpha_j}{\sqcup_i\alpha_i}=\sum_i \sum_j \braket{\alpha_j}{\alpha_i} \neq \sum_i
\braket{\alpha_i}{\alpha_i}}, 
  \end{equation} 
 with the non-vanishing  cross terms encoding the pairwise interference of events.

The quantum vector measure can be constructed from a  strongly positive decoherence functional $\bD: \fS \times \fS
\rightarrow \re^+$  
where $\cH$ is the histories Hilbert space of \cite{hilbert}, with inner product
$\braket{\alpha}{\beta} =D(\alpha,\beta)$.  In this work we will not use the decoherence functional explicitly, {but 
refer the reader to the constructions in \cite{hilbert} and \cite{djs}.}

Since the growth process generates cylinder sets, as in the classical case, we start with defining a vector measure
$\muv$ on $\fZ$, which must at the very least satisfy the analogues of conditions (a), (b) and (c) discussed in Section
\ref{csg.sec}. Since $\fZ$ is {closed under finite set operations} and $\muv$ is additive, we need consider only
the measure on cylinder sets $\cZ$. For any $\cyl{c_n^i} \in \cZ$ we denote the associated state $ \ket{c_n^i} {\in \cH}$ labeled
by the node $c_n^i $ in $\cP$.   

Condition (a) is straightforward to implement since it requires that $\ket{c_n^i}=\ket{c_n^j}$ whenever $c_n^i\sim
c_n^j$, i.e., they 
are order-isomorphic.

{For  condition (b)  we need to use the appropriate analogue of the total probability summing to 1.  Because we want to
construct a  Markovian quantum process on $\cP$,  the  vector measure of a node should be  related to that of its parent
node via a linear transformation on $\cH$. 
Thus for every  child $c_{n+1}^{j(i)}$ of $c_n^i$ we require that  there exists a {\sl transition matrix} $\hO(c_n^i
\rightarrow c_{n+1}^{j(i)})$ such that  
\begin{equation}
\ket{c_{n+1}^{j(i)}}= \hO(c_n^i \rightarrow c_{n+1}^{j(i)}) \ket{c_n^i}. 
  \end{equation} 
Since $\muv$ is finitely additive on $\fZ$, 
\begin{equation}
\label{msr.eq} 
  {\cyl{c_n^i}  =\bigsqcup_{j(i)} \cyl{c_{n+1}^{j(i)}} \Rightarrow}  \sum_{j(i)} \hO(c_n^i \rightarrow c_{n+1}^{j(i)}) = \one
  \end{equation} 
 where $j(i) $ is valued in $\ind(i,n)$ and $\one$ denotes the
 identity operator on $\cH$.  }

What is much more subtle to implement,  is condition (c).  Setting aside the conceptual challenges in
implementing  quantum non-locality, i.e.,  the Bell inequalities \cite{joebell}, even the  straightforward
implementation of spectator independence poses a
challenge in general. However, when $\cH \simeq \bC$ condition (c) or its product form can be unambiguously
implemented, since the transition operators  simplify to   {\sl transition
  amplitudes}  valued in $\bC$.  

{It is relatively straightforward to show that arguments of \cite{Rideout:1999ub} generalises to this complex case, so
that again, the complex growth models can be   characterised in terms of the $\{q_n \} $ or the $\{t_n\}$, with
$q_n, t_n \in \bC$, {where the transition amplitudes $A(c_n^i \rightarrow c_{n+1}^{j_1})$ are given by
\begin{equation}
A(c_n^i \rightarrow c_{n+1}^{j_1})=\frac{\lambda(\varpi,m)}{\lambda(n,0)}, \quad \mathrm{ where} \quad \lambda(a,b)=\sum \limits_{k=b}^{a}
\binom{a-b}{k-b} t_k. 
\label{alphan.eq}
\end{equation}
The quantum measure $\muv(\cyl{c_n^i})$ is then given by  
\begin{equation}\label{eq230201}
\muv(c_n^i)\equiv\ket{c_n^i}=\prod A(c_m \rightarrow c_{m+1}),
\end{equation}
where the product is over transitions along the $(n+1)$-jointed branch of $\cP$ connecting $c_1=e_0$  to the node $c_n^i$.}
We refer to this class of quantum measures as {\sl complex sequential growth} ($\csg$)  models.}

\subsection{Extension of  Complex Measures on $\fZ$}
\label{extension.sec}

The quantum measure space we begin with  is $(\Omega, \fZ,\muv)$, where $\muv$ is constructed from the complex constants
$\{t_0, \ldots, t_n, \ldots \}$, given by Eqn.~(\ref{alphan.eq}) and (\ref{eq230201}).  As in the classical
case, the {measure of an arbitrary covariant event is defined} only if the measure extends to $\fSZ$. However, {while the extension of any probability
  measure on $\fZ$ to $\fSZ$ is guaranteed by Kolmogorov's extension theorem \cite{Kolmogorov:1975}, the extension of a
  vector measure $\muv$ from $\fZ$ to $\fSZ$ exists only if $\muv$ satisfies the conditions of the
  Caratheodary-Hahn-Kluvnek (CHK) extension theorem \cite{diesteluhl}. Importantly, not every $\muv$ given by
  Eq. \ref{eq230201} can be extended to $\fSZ$. }

The convergence condition most relevant to complex measures is that of {\sl bounded variation}.
The  {\sl variation} of $\muv$ is defined as 
\begin{equation}
\var{\muv}(\alpha) \equiv \sup_\pi \sum_{\alpha_i \in \pi} \norm{\ket{(\alpha_i)}}, \quad \forall \,\, \alpha \in \fA
\end{equation}
where $\pi$ is a {\sl finite partition}  of $\alpha$, i.e., $\pi=\{\alpha_1, \ldots, \alpha_k \}, k < \infty
$, $\alpha_i\cap \alpha_j = \emptyset, \forall \,\, i \neq j$ and    $\alpha=\bigsqcup_{i=1}^k \alpha_i$.  The
 measure 
is said to be of {\sl bounded variation}  if 
\begin{equation} 
  \var{\muv}(\Omega) < \infty. 
\end{equation}

In Appendix \ref{vm.app}, we put together existing results in the literature, to show that the  CHK extension theorem for the
complex measure space $(\Omega,\fZ, \muv)$ of interest to us simplifies to the following statement:
\begin{theorem} 
For a complex measure space $(\Omega,\fZ, \muv)$, where $\fZ$ is the event algebra generated from finite set operations on cylinder
sets  and $\muv: \fZ \rightarrow \bC$,  
$\muv$ has a unique extension to $\fSZ$ iff  it is of bounded variation.
\label{complexCHK.thm} 
  \end{theorem} 

Thus bounded variation is  both a necessary and a sufficient
condition for complex measures to extend to $\fSZ$.

In \cite{djs} it was shown that {\sl complex percolation}  ($\bC$P) is of bounded variation iff it is real and
non-negative i.e., $q \in [0,1]$\footnote{Real non-negative $\bC$P is however not a classical measure since the  quantum measure is the
norm or $||\muv(\alpha)||= A(\alpha)^2$, which is non-additive.}.      
The proof makes crucial use of the  Markovian sum rule Eqn.~(\ref{msr.eq}). If $A(c_n^i \rightarrow c_{n+1}^{j(i)}) \in \bC$ denotes
the {\sl transition amplitude} (which is a special case of the transtion matrix of  Eqn.~(\ref{msr.eq})), 
\begin{equation}
\sum_{j(i)} |A(c_n^i \rightarrow c_{n+1}^{j(i)})| \geq 1  \Rightarrow \sum_{j(i)} |A(c_n^i \rightarrow
c_{n+1}^{j(i)})|=1 + \zin, \, \zin \geq 0.
\label{amp.eq} 
\end{equation} 
This inequality is saturated ($\zin =0$) iff the $A(c_n^i \rightarrow c_{n+1}^{j(i)})$  are 
colinear in $\bC$ for all $j(i) \in \ind(i,n)$. 

Since the cylinder sets generate $\fZ$, the boundedness (or lack thereof) of the total variation of $
\Omega$ can be characterised completely by the convergence properties of the constants $\zin$, as one goes
to finer partitions.   At 
every stage $n$, the finiteness of $\Omega_n$ allows one to define 
\begin{equation}
  \zmx:=\max_{c_n^i \in \Omega_n}  \zin, \quad \zmn:=\min_{c_n^i \in \Omega_n}
  \zin.
  \label{zetamxmn.eq}
\end{equation}

As we will see in the following section, these constants can be used to give criteria for bounded variation.

{\section{Extension of the quantum measure in $\bC$SG}
\label{results.sec} 
We present our new results in this section.

Our first result, Theorem \ref{one.thm},  gives  a sufficiency condition for bounded variation of a complex measure
$\muv$ on $\fZ$,   and another
for determining  when it is not, in terms of the constants $\zmx$ and $  \zmn$.

    Subsequently, we  show in Lemma \ref{amax.lemma} and \ref{amin.lemma} that  $\zmx$ and $  \zmn$ are determined
    entirely by transitions from the $n$-antichain $c_n^a$ and $n$-chain $c_n^c$ nodes, respectively. In Eqn.~(\ref{zetaca.eq}) we express $\zmx$ and $  \zmn$ in
    terms of the $\bC$SG  constants $t_n$, which gives us a useful Corollary 
    to Theorem
    \ref{one.thm}.  We then find a large class of non-trivial examples of models in which $\muv$ admits an extension to
    $\fSZ$ as also classes in which such an extension is not possible.

    {  \subsection{Criteria for Bounded Variation}}
  \label{criteria.sec}
    \begin{theorem}\label{one.thm} $\muv$ is of bounded variation if  $\,\,\sum_{n=1}^{\infty} \zmx$ converges. $\muv$ is \textit{not} of bounded variation if $\,\,\sum_{n=1}^{\infty} \zmn$ diverges. \end{theorem}

We find it useful to parse the proof into a set of smaller results.

We start by noting that for any integer $n>0$, $\cZ_n$ forms a partition of $\Omega$, $\Omega=\bigsqcup_{i=1}^{\card_n}
\cyl{c_n^i}$, and therefore by finite additivity we have $\vm{\Omega}=\sum_{i=1}^{\card_n}\vm{c_n^i}$. Define
\begin{equation}
S_n \equiv \sum_{i=1}^{\card_n} \norm{\, \vm{c_n^i} \,}.  
\end{equation}
Since $\norm{ \, \vm{\Omega}\,} =1 $,  $S_n\geq 1$.

\begin{claim}
  $S_n$ is a non-decreasing function of $n$ and satisfies the  inequalities
  \begin{equation} {\prod}_{r=1}^{n-1} (1+\zmnr) \leq S_{n} \leq {\prod}_{r=1}^{n-1} (1+\zmxr).
\label{snzeta.eq} 
\end{equation}
{Therefore, $(i)$
$\lim_{n\rightarrow \infty} S_n<\infty$ if  $\,\, {\sum}_{r=1}^\infty\zeta_r^{\mx}<\infty$, and $(ii)$
$\lim_{n\rightarrow \infty} S_n \rightarrow \infty$ if $\,\,\sum_{r=1}^\infty\zeta_r^{\mn} \rightarrow \infty$.}
 \label{snineq.claim} 
\end{claim} 
\noindent {\bf Proof:} 
\begin{eqnarray} 
  S_{n+1} &=&  \sum_{k=1}^{\card_{n+1}} \norm{ \, \vm{c_{n+1}^{k}}\,}  =\sum_{i=1}^{\card_n} \sum_{j(i)} \norm{ \, \vm{c_{n+1}^{j(i)}}\,}  \nonumber \\
          &=& \sum_{i=1}^{\card_n} \sum_{j(i)} |A(c_n^i \rightarrow c_{n+1}^{j(i)})| \norm{ \, \vm{c_{n}^i}\,}  = \sum_i (1+
  \zin)\norm{ \, \vm{c_{n}^i}\,} ,
\end{eqnarray}
where we have relabelled the $k=\{1, \ldots, \card_{n+1}\} $ nodes in the second equality in terms of  the  parent
nodes $i=\{1, \ldots, \card_n \} $,  and $j(i) \in \ind(i,n) $, the  index set of cardinality  $ |\fC(c_n^i)|$ (as in Eqn.~(\ref{defji.eq})). 
  Since $\zmn \leq  \zin \leq \zmx$, we see that 
  \begin{equation}
   (1+ \zmn)  S_n   \leq  S_{n+1}   \leq  (1+\zmx)   S_n.  
 \end{equation}
 This proves that  $S_n$ is a non-decreasing function of $n$. Applying these  inequalities recursively and noting that
 $S_1=1$ gives us Eqn.~(\ref{snzeta.eq}). {Finally, note that for $a_r \geq 0$, $\prod_{r=1}^\infty
   (1+a_r)$, converges iff $\sum_{r=1}^\infty a_r$ converges \cite{Jeffreys:2006}. This completes the proof.}
\hfill $\qed$

\vskip 1cm

 The following inequalities come in handy to prove the next claim. 
\begin{equation}
\vm{c_n^i}=\sum_{j(i)} \vm{c_{n+1}^{j(i)}} \Rightarrow  \norm{\, \vm{c_n^i}\,} \leq \sum_{j(i)} \norm{\, \vm{c_{n+1}^{j(i)}}
    \,}, 
\end{equation}
for a node $c_n^i$ and its children $\fC(c_n^i)=\{c_{n+1}^{j(i)}\}$.    
  Because of the nesting property of cylinder sets, moreover, for any $m>n$,
 \begin{equation} \cyl{c_n^i} =\bigsqcup_{j(i,m)} \cyl{c_m^{j(i,m)}} \Rightarrow \norm{\, \vm{c_n^i}\,} \leq \sum_{j(i,m)} \norm{\, \vm{c_{m}^{j(i,m)}}
     \,},
   \label{mandn.eq} 
  \end{equation} 
  where $j(i,m)$ takes values in $\ind(i,n,m)$, which label the set of $m$-element descendants of $c_n^i$. (In this notation, $\ind(i,n)=\ind(i,n,n+1)$.) 
\begin{claim}
  $|\muv|(\Omega)=\sup_n S_n$.
  \label{sn.claim} 
\end{claim}
\noindent {\bf Proof:}
Consider any finite partition $\pi$ of $\Omega$. For each $\alpha \in \pi$ consider its fine partition $\cN_{\alpha}$ into $n_\alpha$-element nodes in $\cP$ so that
$\alpha=\bigsqcup_{k\in S_\alpha} \cyl{c_{n_\alpha}^{k}}$.  Then from Eqn.~(\ref{mandn.eq}) 
\begin{equation}
\norm{\, \vm{\alpha} \,}\leq \sum_{k\in S_\alpha}
\norm{\, \vm{c_{n_\alpha}^{k}} \,}. 
\end{equation}
Moreover if $m$ is the largest of the $n_\alpha$ for the partition $\pi$,  for any $\alpha$ with  $n_\alpha <m$ we have
the additional inequality    
\begin{equation} 
\norm{\, \vm{\alpha} \,}\leq \sum_{k\in S_\alpha}
\norm{\, \vm{c_{n_\alpha}^{k}} \,}\leq \sum_{k\in S_\alpha}\sum_{j(k,m)}
  \norm{\, \vm{c_{m}^{j(k,m)}} \,}.
  \end{equation} 
  Since $\{ \cyl{c_{m}^{j(k,m)}}\}$ is an  $m$-level cylinder set
  partition of $\alpha$ for each $\alpha \in \pi$, the union of these partitions provides an $m$-level cylinder set   partition $\cZ_m$ of $\Omega$, so that 
  \begin{equation}
   \norm{\, \vm{\Omega} \,} =1 \leq \sum_{\alpha\in \pi} \norm{\, \vm{\alpha} \,} \leq  \sum_{j=1}^{\card_m} \norm{\,
     \vm{c_m^j} \,} = S_m.
 \end{equation}
 In other words, for any partition $\pi$ of $\Omega$ there exists an $m$ such that
 \begin{equation}
S_m \geq \sum_{\alpha\in \pi} \norm{\, \vm{\alpha} \,} . 
\end{equation}
Since, $\cZ_m$ is itself a partition of $\Omega$, $|\muv|(\Omega) \geq S_m$, {for every integer $m$}.  
This proves the claim. \hfill $\qed$ 
\vskip 1cm
\noindent {\bf Proof to theorem \ref{one.thm}}: Since from Claim \ref{sn.claim}  the variation of $\muv$ depends
only on the $S_n$, along with Claim \ref{snineq.claim}, this completes the proof. 
\hfill $\qed$

\subsection{ Criteria for Bounded Variation in $\bC$SG}

\label{complex.sec}  

We now translate the convergence criterion Theorem \ref{one.thm} to requirements on the coupling constants $t_n$
for $\csg$. We find the important result that transitions from the $n$-antichain node $c_n^a$ determines
$\zeta_n^{\mx}$  while the $n$-chain node $c_n^c$ determines $\zeta_n^{\mn}$. This gives an explicit functional form for
$\zeta_n^{\mx}, \zeta_n^{\mn}$ in terms of the $t_n$.

Let us first define some notation. Consider the set of possible transitions from a node $c_n^j$ and let  $\cT(c_n^j)$ denote the list of the (possibly repeated) $(\varpi, m)$ values for these transitions. Then by the Markov sum rule,   
\begin{equation}
\sum_{(\varpi, m) \in \cT(c_n^j)} \frac{\lambda(\varpi,m)}{\lambda(n,0)} = 1  \Rightarrow  \zeta_n^j=\sum_{(\varpi, m) \in
  \cT(c_n^j)} \frac{|\lambda(\varpi,m)|}{|\lambda(n,0)|} -  1 \geq 0.
\label{snj.eq} 
  \end{equation} 

For  $m < n$ we say that $c_{m}^{k}$ is a  {\sl partial stem}  in
$c_n^j$ if  (i) $c_{m}^{k} \subset
c_n^j$  and (ii) for all $e \in c_{m}^{k} $, $\mathrm{past}(e)
\subseteq c_{m}^{k} $.  Let $P_m(c_n^j)$ denote the set of all $m$-element partial stems in
$c_n^j$.  For  $m=n-1$, we note that the parent node of $c_n^j$ in $\cP$ is one
of the partial stems in $P_{n-1}(c_n^j)$.  While the rest of the partial stems in $P_{n-1}(c_n^j)$ 
are each  order-isomorphic to some $(n-1)$-element node in $\cP$ they are {\it not} themselves nodes, since they are not naturally labelled. 
Moreover, every partial stem $c_{n-1}^k\in P_{n-1}(c_n^j)$, is associated with a unique  element $e_s\equiv c_n^j
\backslash c_{n-1}^k$  which must be maximal in $c_n^j$. 

For any given $c_{n-1}^k \in P_{n-1}(c_n^j)$, we  can therefore parse the transitions from $c_n^j$ into (A) the  set of transitions which only involve $c_{n-1}^k$, so that  $e_s= c_n^j\backslash c_{n-1}^k$ is always in the  spectator set, plus (B) the set of transitions that always include $e_s $
in the precursor set. In doing this one can relate transition amplitudes from $c_n^j$ to  those from $c_{n-1}^k$. Let
$l_{A}(j)$ label the type (A) children of $c_n^j$, and similarly let $l_{B}(j)$ label the type (B) children of $c_n^j$.
For any  transition of
type (A), $c_n^j\rightarrow c_{n+1}^{l_A(j)}$, there exists a child  $c_{n}^{j(k)}$ of $c_{n-1}^k$ (where $j(k) \in \ind(k,n-1)$) such that
$c_{n}^{j(k)} \sim c_{n+1}^{l_A(j)}\backslash e_s$. This allows us to re-express the transition amplitude as 
\begin{equation}
A(c_n^j\rightarrow c_{n+1}^{l_A(j)})=A(c_{n-1}^{k} \rightarrow c_{n}^{j(k)}) \times
\frac{\lambda(n-1,0)}{\lambda(n,0)}. 
\end{equation}
Summing over all the transitions from $c_{n}^{j}$, for the given choice of partial stem $c_{n-1}^k$ we find 
\begin{eqnarray} 
  \sum_{l(j)} A(c_{n}^{j} \rightarrow c_{n+1}^{l(j)}) \!\!\!& =& \!\!\! \sum_A
                                                              A(c_{n}^{j} \rightarrow c_{n+1}^{l_A(j)}) + \sum_B
                                                              A(c_{n}^{j} \rightarrow c_{n+1}^{l_B(j)}), \nonumber \\ 
                                                           \!\!\!&=& \!\!\! \biggl(\sum_{i(k)} A(c_{n-1}^{k} \rightarrow
                                                                c_{n}^{i(k)})\biggr) 
                                                                \frac{\lambda(n-1,0)}{\lambda(n,0)} +\!\!  \sum_B
                                                                     A(c_{n}^{j} \rightarrow c_{n+1}^{l_B(j)}),
                                                                     \label{psone.eq} 
\end{eqnarray} 
where  $l(j) \in \ind(j,n)$, and $i(k) \in \ind(k,n-1)$. Applying the Markov sum rule to the LHS as well as the term in brackets we see that 
\begin{equation} 
\sum_B A(c_{n}^{j} \rightarrow c_{n+1}^{l_B(j)}) =  \frac{\lambda(n,1)}{\lambda(n,0)}  
\Rightarrow  \sum_B |A(c_{n}^{j} \rightarrow c_{n+1}^{l_B(j)})|  \geq    \frac{|\lambda(n,1)|}{|\lambda(n,0)|},  \label{psfour.eq} 
\end{equation}

Defining $Q_n^j \equiv \zeta_n^j+1 \geq 0$, Eqn.~(\ref{psone.eq}) and (\ref{psfour.eq}) give the useful identities 
\begin{eqnarray} 
   Q_n^j &=&Q_{n-1}^{i(j)} \frac{|\lambda(n-1,0)|}{|\lambda(n,0)|} + \sum_B |A(c_{n}^{j} \rightarrow
                   c_{n+1}^{l_B(j)})| \label{snjone.eq}  \\ 
  \Rightarrow  Q_n^j &\geq&  Q_{n-1}^{i(j)} \frac{|\lambda(n-1,0)|}{|\lambda(n,0)|} + \frac{|\lambda(n,1)|}{|\lambda(n,0)|}. \label{snjtwo.eq} 
\end{eqnarray}

For the $n$-antichain  node $c_n^a$, for each transition, $m=\varpi$,  i.e., the number of maximal elements is equal
to the cardinality of the precursor set. Hence
\begin{equation}
A(c_n^a \rightarrow c_{n+1}^{j(a)})=\frac{t_m}{\lambda(n,0)}, 
\end{equation}
where $j(a)$ labels the set of children of $c_n^a$.   
For fixed $m$ there are $\binom{n}{m}$ possible choices of precursor sets for the new  element $e_{n+1}$. Hence 
\begin{equation}
 Q_n^a=\frac{\sum_{k=0}^{n}\binom{n}{k}|t_k|}{|\lambda(n,0)|} 
 \label{sna.eq} 
\end{equation}
Inserting this into Eqn.~(\ref{snjone.eq}) we find that for the antichain
{\begin{equation}
 \sum_B |A(c_{n}^a \rightarrow
                   c_{n+1}^{l_B(a)})|= \frac{\sum_{k=1}^n\binom{n-1}{k-1} |t_k|}{|\lambda(n,0)|} \geq \frac{|\lambda(n,1)|
                   }{|\lambda(n,0)|},  
                 \end{equation}}

where $l_B(a)$ labels the set of type (B) children of $c_n^a$,  so that  
{\begin{equation}
  Q_n^a = Q_{n-1}^{a} \frac{|\lambda(n-1,0)|}{|\lambda(n,0)|} +  \frac{\sum_{k=1}^n\binom{n-1}{k-1} |t_k|}{|\lambda(n,0)|}. 
\label{sna.eqn}
\end{equation}}

For the $n$-chain node  $c_n^c$, there is a  unique $(n-1)$-element  partial stem, the $(n-1)$-chain $c_{n-1}^c$, with 
$e_s=e_n$. For this node,  the only possible transition of  type (B)  is that  with $e_n$ as the (unique) maximal element of the precursor set, 
i.e., $c_n^c \rightarrow c_{n+1}^c$. In this case, Eqn.~(\ref{snjone.eq})  reduces to 
\begin{equation}
  Q_n^c = Q_{n-1}^{c} \frac{|\lambda(n-1,0)|}{|\lambda(n,0)|} +  \frac{|\lambda(n,1)|}{|\lambda(n,0)|}.
  \label{snc.eq}
\end{equation}

We are now equipped to prove the main results of this section.

\begin{lemma}\label{amax.lemma}
$\zeta_n^{max}=\zeta_{n}^a$.
\end{lemma}
\noindent {\bf Proof:}
For any  node $c_n^j$ 
\begin{equation} 
  \sum_{(\varpi, m) \in \cT(c_n^j)} \lambda(\varpi,m) = \lambda(n,0) = \sum_{k=0}^n\binom{n}{k} t_k.
\end{equation}
$\cT(c_n^j)$ therefore provides a node dependent partition of $\lambda(n,0)$, with $\cT(c_n^a)$ being  the finest such
partition, given by the second equality.  Since  $Q_n^j=\sum_{(\varpi, m) \in \cT(c_n^j)} |\lambda(\varpi,m)|$ and
$Q_n^a=\sum_{k=0}^n\binom{n}{k} |t_k|$, this means that  $Q_n^a \geq Q_n^j $. 
\hfill $\qed$

\begin{lemma}\label{amin.lemma}
$\zeta_n^{min}=\zeta_{n}^c$.
\end{lemma}
\noindent {\bf Proof:} We prove this inductively. For $n=1,2$ we see that 
\begin{eqnarray} 
Q_1^{a,c} & = & \frac{|t_0|+|t_1|}{|t_0+t_1|}=1+\zeta_1   \nonumber \\ 
 \Rightarrow Q_2^c =  \frac{|\lambda(1,0)|}{|\lambda(2,0)|} (1+\zeta_1)+ \frac{|\lambda(2,1)|}{|\lambda(2,0)|}, &&  
     Q_2^a =  \frac{|\lambda(1,0)|}{|\lambda(2,0)|} (1+\zeta_1)+ \frac{\sum_{k=1}^2\binom{1}{k-1}
                |t_k|}{|\lambda(2,0)|} \nonumber \\
              \Rightarrow    Q_2^a &\geq&  Q_2^c.
\end{eqnarray} 
Now, assume that $Q_{n-1}^j \geq Q_{n-1}^c$ for all $j\in \ind(n)$, where $\ind(n)=\{1,\ldots, \card_n\}$ as before.  Then from Eqn.~(\ref{snjtwo.eq}) and Eqn.~(\ref{snc.eq})
\begin{equation}
  Q_n^j \geq   Q_{n-1}^c \frac{|\lambda(n-1,0)|}{|\lambda(n-1,0)|} + \frac{|\lambda(n,1)|}{|\lambda(n,0)|} =Q_n^c, 
\end{equation}
which proves the claim. 
\hfill $\qed$

Eqn.~(\ref{snc.eq}) and (\ref{sna.eq}) also implies  
\begin{equation}
\zeta_n^a=\frac{\sum_{k=0}^{n}\binom{n}{k}|t_k|}{|\lambda(n,0)|} -1, \quad
\zeta_n^c=\frac{\sum_{\varpi=1}^n|\lambda(\varpi,1)|}{|\lambda(n,0)|} + \frac{|\lambda(0,0)|}{|\lambda(n,0)|}-1. 
\label{zetaca.eq}
\end{equation}

Putting this together with Theorem \ref{one.thm} we have the result
\begin{corollary}
For the $\bC$SG dynamics $\muv$ is of bounded variation if  $U_a\equiv \,\,\sum_{n=1}^\infty \zeta_n^a$ converges and $\muv$ is not of
bounded variation if $U_c\equiv \,\,\sum_{n=1}^\infty \zeta_n^c$ does not
converge, where $\zeta_n^a,\zeta_n^c$ are given by Eqn.~(\ref{zetaca.eq}).
\label{zetaca.cor} 
  \end{corollary}

  \subsection{ Existence and Non-Trivial Examples}
\label{examples.sec}

  From Eqn.~(\ref{zetaca.eq}) it is clear that $\zeta_n^{c,a}=0$ for all $n$ iff the $t_k$ are all colinear. Since
  $t_0=1$ this means that the $t_k$ must all lie on $\re^+$. For such  $\bC$SG or $\re^+$SG dynamics,   convergence is trivially
  satisfied, so that we have 
  \begin{corollary}
For  $\re^+$SG dynamics (i.e.,  with all $t_k \in \re^+$)   $\muv$ is of bounded variation. 
    \end{corollary} 

While this establishes the existence of covariant $\bC$SG dynamics, $\re^+$SG is too restricted a subclass and it is
therefore of interest to look for non-trivial examples of  complex covariant dynamics, i.e., with non-vanishing phases.

We compare  $U_a$ and $U_c$ (defined in Corollary \ref{zetaca.cor}) term by term with the series $U_x\equiv \sum_{n=1}^\infty \frac{1}{n^x}$, which
converges for $x>1$ and diverges otherwise.   Thus, our requirement for convergence of $U_a$  is that there exists an $n_0<\infty$ and an $x>1$, such that for
all $n>n_0$,   $\zeta_n^a < \frac{1}{n^x}$. This means that the complex measure extends. Conversely, if for any $x>1$,  there exists an $n_0<\infty$ such that   $\zeta_n^c >  \frac{1}{n^x}$ for all $n>n_0$, then $U_c$ diverges. This means that the complex measure does not extend. 
It will be useful to define the expression  
\begin{equation}
L_n^{a,c}(x) \equiv  \zeta_n^{a,c}-\frac{1}{n^x}.
\end{equation}
to check for convergence or divergence. 

\subsubsection{Finite number of non-zero couplings}

The simplest non-trivial case  is $t_k\neq 0$ for some $k>0$ and $ t_{k'}=0, \forall k'\neq k, k'>0$.  Let $t_k=s e ^{i\phi}$, $s\in \re^+$. Then
\begin{equation}
  \zeta_n^a=\frac{1+R_k(n)s}{\sqrt{1+2s R_k(n) \cos\phi+ s^2 R_k(n)^2}}-1, 
\end{equation}
where we use the shortform $R_k(n)\equiv \binom{n}{k}$.

We now look for conditions on $s,k$ and $\phi$ such that $L_n^a(x) < 0$ for large $n$ and $x>1$. 
Since $\zeta_n^a \geq 0$,   $L_n^a(x) < 0$ implies that 
\begin{equation} 
 \biggl( -\frac{2}{n^x}-\frac{1}{n^{2x}}\biggr)\biggl(1+s^2 R_k(n)^2\biggr)+2s R_k(n) \biggl((1-\cos\phi) -\biggl(
 -\frac{2}{n^x}-\frac{1}{n^{2x}}\biggr)\cos\phi \biggr) <  0.
 \label{conv.eq} 
\end{equation}

For  $n>>k$ we can use the asymptotic form $\binom{n}{k} \sim \frac{n^k}{k!}$  
to show that the dominant contribution to the LHS is   
\begin{equation}
\approx \frac{2s}{ k!}   n^k   \biggl(-\frac{s}{k!} n^{k-x} + (1-\cos\phi)\biggr). 
\end{equation}
For this to be negative in the large  $n$ limit, the first term  must dominate, or $k>x>1$, with no restrictions on $s,\phi$.  Thus, we see that the measure is of bounded variation for all choices of $t_k \in \bC$ as long as  $k\geq 2$.   

When $k=1$,
\begin{equation}
{  \zeta_n^c=\zeta_n^a=\frac{ns+1}{\sqrt{1+n^2s^2+2ns\cos(\phi)}}-1 = \frac{1}{ns} + O\biggl(\frac{1}{n^2s^2}\biggr),} 
\end{equation}
which means that the measure is not of bounded variation. 
 
This simple example can be easily generalised to include an arbitrary but finite number of couplings.  
 
Let $\{t_0, t_{k_1}, t_{k_2} , \ldots t_{k_m}\}$ be a finite set of non-zero  coupling constants where wlog we take  $k_m >  k_{m-1} \ldots >  k_1 >0$. Let
$t_{k_i}=s_ie^{i \phi_i}$, $s_i\in \re^+$ and $R_i=\binom{n}{k_i}$. Then
\begin{equation}
\zeta_n^a=\frac{1+ \sum_{i=1}^m R_i s_i}{|1+ \sum_{i=1}^m R_i s_i e^{i\phi_i}|} -1 
  \end{equation} 
Requiring that $\zeta_n^a< \frac{1}{n^x}$ for some $x>1$ leads to the inequality  
\begin{eqnarray} 
 \biggl( -\frac{2}{n^x}-\frac{1}{n^{2x}}\biggr)\biggl(1+ \sum_{i}R_i^2s_i^2\biggr) + 2
 \sum_iR_is_i\biggl(1-\cos \phi_i + \biggl( -\frac{2}{n^x}-\frac{1}{n^{2x}}\biggr) \cos\phi_i\biggr)  &&\nonumber
  \\ + 2\sum_{i,j, i\neq j}R_iR_js_is_j
 \biggl( 1-\cos(\phi_i-\phi_j) +  \biggl( -\frac{2}{n^x}-\frac{1}{n^{2x}}\biggr) \cos(\phi_i-\phi_j) \biggr) < 0. && 
\label{genconv.eq}
 \end{eqnarray} 
For $m>1$  the dominant contributions to the LHS for large $n$,  arising from the  $k_m $
and $k_{m-1}$ terms are  
  \begin{equation}
    -\frac{2 s_m^2}{(k_m!)^2}n^{2k_m-x} + \frac{2s_ms_{m-1}}{k_m! k_{m-1}!} n^{k_m+k_{m-1}}(1-\cos(\phi_m-\phi_{m-1})).
    \label{finitem.eq} 
  \end{equation}
  For  this to be negative, $2k_m-x > k_m+{k_{m-1}} \Rightarrow k_m-k_{m-1} > {x}$, {which implies bounded variation whenever $k_m-k_{m-1} >1$,} with no  restrictions on the $s_i, \phi_i$.
  
  On the other hand, if $k_{m}-k_{m-1}=1$, then the second term in Eqn.~(\ref{finitem.eq}) dominates
  which means that $L_n^a(x)>0$. Unlike the $m=1$ case, however this  is not sufficient to prove
  divergence.

 Combining these results we have proved the following
  \begin{claim}
    Let  $\{t_0, t_{k_1}, \ldots ,t_{k_m} \}$ be  the only non-zero $\bC$SG coupling constants.

\noindent The $\bC$SG dynamics  is of bounded variation if any one of the following is true
\begin{enumerate}
\item $t_{k_i}\in \re^+, i \in \{0,\ldots, m\}$.
  \item $m=1$ and $k_1>1$.  
  \item  $1<m<\infty$, $k_m-k_{m-1}>1$.
\end{enumerate}
\noindent It is not of bounded variation if $t_1\not\in \re^+$  and $m=1$, $k_1=1$. 

\label{finite.claim}
\end{claim}

\subsubsection{Countable number of non-zero couplings}

For a countable  number of couplings we cannot use the above approximations, and we turn to more general arguments to show existence for non-real $t_k$. 

The criterion for convergence is roughly that that the $\zeta_n^a$   become sufficiently small as $n$ increases. This in turn  means
that  the amplitudes in the  transition  at stage $n$ become increasingly colinear according to
Eqn.~(\ref{amp.eq}). 

Let us examine this using  an  explicit example. Consider a set of countable couplings such that for  $k>k_0>0$,  $t_k=s_ke^{i\phi_0}$, i.e., the $t_k$ become  colinear for $k>k_0>0$. Then we can express 
\begin{equation}
\zeta_n^a = \frac{\sum_{k<k_0} \binom{n}{k} |t_k| + |I_0^n|}{|\sum_{k<k_0} \binom{n}{k} t_k +I_0^n|} -1,  
\end{equation}
where $I_0^n\equiv \sum_{k>k_0}^{\infty} \binom{n}{k} t_k= e^{i\phi_0}
\sum_{k>k_0}^{\infty} \binom{n}{k} s_k$, so that $|I_0^n| \equiv \sum_{k>k_0}^{\infty} \binom{n}{k}
s_k$. 

As in the finite coupling case, the requirement that 
$\zeta_n^a<\frac{1}{n^x}$ for all $  x>1$ simplifies to 
\begin{eqnarray} 
 \biggl( -\frac{2}{n^x}-\frac{1}{n^{2x}}\biggr)\biggl(\sum_{i=0}^{k_0}R_i^2s_i^2 +|I_o^n|^2\biggr)
 & &  \nonumber
  \\ + {2}\sum_{i,j, i\neq j}^{k_0}R_iR_js_is_j
 \biggl( 1-\cos(\phi_i-\phi_j) +  \biggl( -\frac{2}{n^x}-\frac{1}{n^{2x}}\biggr) \cos(\phi_i-\phi_j) \biggr)  && \nonumber
  \\ + 2\sum_{i=0}^{k_0}R_is_i|I_0^n|
 \biggl( 1-\cos(\phi_i-\phi_0) +  \biggl( -\frac{2}{n^x}-\frac{1}{n^{2x}}\biggr) \cos(\phi_i-\phi_0) \biggr) <0
 \end{eqnarray} 
The largest possible contribution  from the  $R_i$ goes like $\frac{n^{k_0}}{k_0!}$. If $s_k$ is a
growing function of $k$, then 
$|I_0^n| $ grows at least as fast as $ \sim \binom{n}{\frac{n}{2}} s_{\frac{n}{2}}\sim 2^{n-1}
s_{\frac{n}{2}}$  and hence dominates the contribution from the $R_i$. Thus the dominant contribution to the LHS is 
\begin{equation}
\approx -2 |I_0^n|^2n^{-x} + \frac{2}{k_0!} n^{k_0}|I_0^n|s_{k_0}(1-\cos(\phi_{k_0}-\phi_0)).
  \end{equation}
  This is negative for large $n$ if
  \begin{equation}
    |I_0^n| > n^{k_0+x}.
    \label{inot.eq} 
    \end{equation} 
Let us consider a couple of  specific examples. (i)  $s_k=s^k, k>k_0$, for any $s$, since for large enough $n$,
$|I_0^n|\approx  (1+s)^n$ which clearly satisfies this condition. (ii) $s_k=2^{2k}$, for which
$|I_0^n|\approx 2^{2n}$.  

We have thus  shown that 
\begin{claim} 
  The complex measure of the $\bC$SG dynamics given  by the countable  set of coupling constants
\begin{equation} 
  \{t_0, t_1, \ldots t_{k_0},
  s_{k_0+1}e^{i\phi_0}, s_{k_0+2}e^{i\phi_0}, \ldots s_k e^{i\phi_0}, \ldots \}
  \end{equation} 
is extendible for $k_0<\infty$ for  $s > 0$ and  (i) $s_k=s^k$ or  (ii) $s_k=2^{2k}$.  
\label{countable.claim} 
\end{claim} 
 
Our analysis makes it possible to find other,  less simplistic, dynamics for which the complex
measure extends to $\fSZ$ , but we will not explore these further in this work.  
 
The example of   ($\bC$P)  examined in   \cite{djs}, on the other hand, does {\it not} satisfy this
asymptotic  colinearity condition for $0<\phi<2\pi$ since $t_k=t^k=s^ke^{ik\phi}$. Thus, as  $k$ increases, the phase does not
stabilise.  We discuss this case briefly using the perspective we have gained in our analysis.

In $\bC$P, $t_k=t^k, q_k=q^k$ and $t=\frac{1-q}{q}$.  Note that $t$ is real and positive if and only if $q$ is real and
$0<q\leq 1$. Using 
\begin{equation}
  \lambda(\varpi, 1)= \frac{1-q}{q^{\varpi}}, \quad \lambda(n, 0)=\frac{1}{q^n}, 
\end{equation}
we see that
\begin{equation}
  \zeta_n^c=|1-q|\sum_{\varpi=1}^n|q|^{n-\varpi} + |q|^n -1.
  \label{cpzeta.eq} 
\end{equation}
For $|q|=1$, $q\neq 1$,
\begin{equation} 
  \zeta_n^c= n\times |1-q|
\end{equation}
and hence the sum $S_c\equiv\sum_{n}^\infty \zeta_n^c$ is explicitly divergent.

If $|q|>1$, the $|q|^n$ term in Eqn.~(\ref{cpzeta.eq}) dominates and again leads to a divergence in the sum $S_c$. 
If  $|q|<1$, $q \not\in \re^+$, 
\begin{equation}
\zeta_n^c=(1-|q|^n)\biggl(\frac{|1-q|}{1-|q|}-1\biggr) \Rightarrow S_c = \biggl(\frac{|1-q|}{1-|q|}-1\biggr)
\sum_{n=1}^\infty (1-|q|^n) 
\end{equation}
which is again divergent

This gives us an alternate proof that $\bC$P is not of bounded variation unless  $q\in[0,1]$.

\section{Discussion}
\label{discussion.sec}

In this work we have shown that the quantum measure extends from the event algebra $\fZ$ to
$\fS_\fZ$ for several classes  of $\bC$SG models. We also find new classes of $\bC$SG models  
in which it does not extend.  Importantly, for the former class of dynamics, this implies that  {\it every}  covariant event in $\fS_\fZ$ is
measurable. Thus, one may attempt to answer physically interesting questions in these models.

The simplest question to ask is whether the dynamics is originary. As  discussed in the
introduction, the originary  event $\alpha_{\mathrm{orig}}$ is the set  of all causal sets for which  there is an element $e_0$ to the past of
all other  elements.  As shown in \cite{Brightwell:2002yu,dz} the  {\sl stem event} associated with every
node $c_n^j$  
\begin{equation}
\stem(c_n^i) = \{c \in \Omega| c_n^i\mathrm{\,\, is \,\, a \,\, partial \,\, stem \,\,in \,\,} c\},  
\end{equation}
is itself covariant and hence belongs to $\fSZ$ but not $\fZ$ . The originary event of Section \ref{gp.sec} is then simply 
{$\alpha_{\mathrm{orig}} = \stem(c_2^a)^c$}, where 
\begin{equation}
  \stem(c_2^a)= \bigsqcup_{{n>0}} \bigsqcup_{i \in \cI_n} \cyl{c_n^i}, 
 \end{equation}  
over all {$n>0$}  and where $\cI_n$ labels the nodes for which the $n^{\mathrm{th}}$ element is the only gregarious one.  
Thus when the measure on $\fZ$ extends to $\fSZ$, 
\begin{equation}
\ket{\mathrm{orig}}=\ket{\Omega}-\ket{\stem(c_2^a)} = \one- \sum_{{n>0}} \sum_{i \in \cI_n} \ket{c_n^i}. 
\end{equation} 
At each stage, the factorisation of the amplitude allows us to express   
\begin{equation}
\sum_{i\in \cI_n}\ket{c_n^i}=\sum_{j\not\in \cI_{n-1}}\ket{c_{n-1}^j} \hq_n=\biggl(\one-\sum_{k=0}^{n-1}\sum_{i_k\in \cI_k}
\ket{c_k^{i_k}}\biggr)\hq_n
  \end{equation} 
  where $\hq_n$ is the amplitude for the gregarious transition. Simplifying we see that
  \begin{equation} 
  \ket{\mathrm{orig}}=\Pi_{i=1}^\infty \biggl(\one- \hq_i\biggr)
\end{equation}
This expression can now be evaluated for each of  the possible extendible $\bC$SG dynamics we have considered.

The evaluation becomes trivial for any  dynamics in which $t_1=0$, since $q_1=1 $. For the class of
$\bC$SG measures that do extend (see  Claims \ref{finite.claim} and \ref{countable.claim})   we 
conclude that $\ket{\mathrm{orig}}=0$ whenever $t_1=0$.  Using the \emph{principal of preclusion} which states that
(covariant) sets of quantum measure zero do not
happen, we see that for this class of  dynamics we can make the somewhat trivial, but predictive
statement that the 
originary event {\it never} happens.  It is expected that such preclusions can also  occur when $q_1\neq 0$, when there
are subtle phase cancellations. We leave such an investigation to future work.

For $\bC$P, which we have seen does {\it not}
extend, the expression on the RHS has the simple form of the Euler Totient function \cite{djs,ss-ec} and is
finite for $|q|\leq 1$. 
We expect that the measure will depend on this function for the class of dynamics which converges to $\bC$P
at larger $k$.  We postpone a detailed analysis  of this to future work, as also explicit calculations of the measure
of other covariant observables.

\vskip 0.5cm

\textbf{Acknowledgments:} {The authors would like to thank Fay Dowker for valuable
discussions.} This research was partly supported by the Stevenson Fund, Imperial College London. SZ thanks
Raman Research Institute for hospitality while this work was being completed. SZ is partially supported by the Kenneth
Lindsay Scholarship Trust. SS  is supported in part by a Visiting Fellowship at the Perimeter Institute. 

\clearpage
\appendix

\appendix
\section{Some Basic Definitions in Causal Set Theory}
\label{defs.app}

This section contains the definitions of various standard terms in CST that have appeared in the preceding sections. 

\begin{itemize}

\item A  causal set \emph{sample space} is a collection of causal sets. For sequential growth, this is the
  collection  $\Omega$ of countable, labelled, \emph{ past
    finite} causal sets, i.e.,
  \begin{equation}
\Omega \equiv \{c| \forall e \in c, |{\mathrm{Past}(e)}|< \infty \} 
    \end{equation}

\item An \emph{event} is a measurable subset of $\Omega$

  \item A \emph{covariant observable} $\cO\subset \Omega$ is a measurable subset of $\Omega$ such that if $c \in \cO$,
    then so is every relabelling of $c$.

 \item  An $n$ element  \emph{chain} is a completely ordered $n$-element set $c$, i.e., for every $e_i, e_j \in c$, either $e_i \prec e_j$ or
   $e_j \prec e_i$.   An $n$-element  \emph{antichain} is a set of mutually unrelated elements: $e_i\not\prec e_j \,\,\forall \,\,e_i, e_j\in c$. 

 \item  \emph{{Poscau}} $\cP$ refers to the tree of labelled causal sets. A \emph{node} in $\cP$ is a finite
   element labelled causal set. 

 \item A \emph{cylinder set} $\cyl{c_n^i} \subseteq \Omega$ such that 
   \begin{equation}
\cyl{c_n^i} \equiv \{ c| c|_n=c_n^i\} 
     \end{equation} 
where $c|_n$ denotes  the first $n$ elements of $c$.   

\end{itemize}

\section{CHK for $\cH\sim \bC$}
\label{vm.app}
{
  We now state the relevant parts of the  Caratheodary-Hahn-Kluvnek theorem\footnote{The theorem as stated in
    \cite{diesteluhl} has two more equivalent conditions but they are not of direct relevance to this work, so we omit
    them.}  \cite{diesteluhl}.
\begin{theorem}
Let $\fA$ be a field of subsets of $\Omega$ and $\fS_\fA$ be the $\sigma$-field generated by $\fA$. Then if $\muv$ is a
(i) {\sl bounded}, (ii) {\sl weakly countably additive}  vector measure over $\fA$ then the following are equivalent.
\begin{enumerate}
\item $\exists \,\,\, ! $ countably additive extension of $\muv$ to  $\fS_{\fA}$.
  \item $\muv$ is (iii) {\sl strongly  additive}.  
  \end{enumerate} 
    \label{chk.thm} 
    \end{theorem} 

    We define the terminology used in the theorem below.
    \begin{enumerate}
      
    \item The {\sl semi-variation} $||\muv||$ of a vector measure $\muv$  is defined as
    \begin{equation}
{||\muv||(\alpha)=sup \{ |x^*\muv|(\alpha) ; x^* \in \cH^*, \norm{x^*} \leq 1 \} , }
      \end{equation}
      where $\cH^*$ is the dual space. Note that {$x^*\muv$} is an inner product measure, itself valued in $\bC$.
      $\muv$  is said to be {\sl bounded} if $||\muv||(\Omega) < \infty$. 
      
    \item If  for every infinite  sequence $ \{\alpha_1, \ldots, \alpha_n, \ldots\}
    $ {of pairwise disjoint members of $\fA$} such that  $\bigcup_{i} \alpha_i \in \fA$,   $\muv (\bigcup_{i} \alpha_i) = \sum_i
    \muv(\alpha_i)$, then $\muv$ is {\sl  countably additive.} 
   \item    $\muv$ is {\sl weakly countably additive} if $x^*\muv$ is { countably additive} {for every $x^* \in  \cH^*$.}
     
\item $\muv$ is {\sl strongly additive}  {if} for every sequence  $\{ \alpha_n\}$ of pairwise disjoint element of $\fA$,
  $\sum_{n=1}^\infty \ket{\alpha_n}$ converges in the norm.   
     \end{enumerate} 
  We now show how the CHK theorem simplifies to  Theorem \ref{complexCHK.thm}.

  \noindent {\bf Proof of Theorem \ref{complexCHK.thm}: }

  From \cite{diesteluhl} if $\muv$ is of  bounded variation, then it is strongly additive, which in turn implies that it
  is bounded.  For $\cH \sim \bC$, the converse can be proved, i.e., boundedness implies bounded variation. Since the former implies that
  $|x^*\muv|(\Omega) < \infty$ for all $x^*\in \cH^*$, by putting $x^*=1$ we see that $|\muv|(\Omega) < \infty$.
  Thus bounded variation is equivalent to the conditions of boundedness and strong additivity. 

  Since $\fA=\fZ$, for every $\alpha \in \fZ$ there exists a smallest $n <\infty$ and a subset $S \subset \{ 1 , \ldots,
  \card_n\}$ such that $\alpha=\bigsqcup_{k\in S} \cyl{c_n^k}$. Thus, $\muv$ is trivially countably and weakly countably
  additive.

  Using the CHK theorem,  this means that bounded variation of $\muv$ is sufficient for it to extend  to $\fS_\fZ$.   

That it is also necessary, comes from Theorem 6.4 in \cite{rudin},  which states that a complex measure on any
$\sigma$-algebra is of bounded variation. This completes the proof. \hfill $\qed$

\bibliography{LSA}
\bibliographystyle{unsrt}

\end{document}